\documentclass[aps,pra,reprint,showpacs,superscriptaddress,longbibliography]{revtex4-1}
\usepackage{mathtools,amssymb,graphicx,units}

\makeatletter
    \renewcommand\@make@capt@title[2]{%
     \@ifx@empty\float@link{\@firstofone}{\expandafter\href\expandafter{\float@link}}%
      {\textbf{#1}}\@caption@fignum@sep#2\quad}%
\makeatother
\makeatletter 
\renewcommand{\fnum@figure}{\textbf{Figure~\thefigure}}
\makeatother

\usepackage{graphicx}

\usepackage[plainpages=false,pdfpagelabels,colorlinks=true,linkcolor=red,urlcolor=blue,citecolor=blue,pdftitle={Title},pdfauthor={},pdfdisplaydoctitle=true,pdfduplex=DuplexFlipLongEdge]{hyperref}

\usepackage{verbatim}
\usepackage[english]{babel}
\usepackage{color}

\usepackage[normalem]{ulem}

\newcommand{\beginsupplement}{%
        \setcounter{table}{0}
        \renewcommand{\thetable}{S\arabic{table}}%
        \setcounter{figure}{0}
        \renewcommand{\thefigure}{S\arabic{figure}}%
}

\begin{document}
\definecolor{darkred}{rgb}{0.80,0,0}
\definecolor{blood}{rgb}{0.50,0,0}
\definecolor{brightred}{rgb}{1,0,0}
\definecolor{orange}{rgb}{1,0.3,0}
\definecolor{bluegreen}{rgb}{0,0.5,0.5}
\definecolor{lightblue}{rgb}{0,0.5,0.8}
\definecolor{darkgreen}{rgb}{0,0.5,0}
\definecolor{green}{rgb}{0,0.70,0}
\definecolor{darkblue}{rgb}{0,0,0.80}
\definecolor{magenta}{rgb}{1,0,1}
\definecolor{softmagenta}{rgb}{0.85,0.1,0.6}
\definecolor{mauve}{rgb}{0.6,0.1,1}
\definecolor{white}{rgb}{1,1,1}
\definecolor{black}{rgb}{0,0,0}

\title{Classical Analog  of Quantum Models in Synthetic Dimensions}
\author{Max Cohen}
\affiliation{Department of Physics and Astronomy, University of California, Davis, CA 95616,USA}
\author{Max Casebolt} 
\affiliation{Department of Physics and Astronomy, University of California, Davis, CA 95616,USA}
\author{Yutan Zhang}
\affiliation{Department of Physics and Astronomy, University of California, Davis, CA 95616,USA}
\author{Kaden 
R. A. Hazzard}\affiliation{Department of Physics and Astronomy, University of California, Davis, CA 95616,USA}
\affiliation{ Department of Physics and Astronomy, Rice University, Houston, TX 77005}
\affiliation{Rice Center for Quantum Materials, Rice University, Houston, TX 77005}
\author{Richard Scalettar}
\affiliation{Department of Physics and Astronomy, University of California, Davis, CA 95616,USA}

\date{\today}

\begin{abstract}
We introduce a classical analog of quantum matter in ultracold molecule- or Rydberg atom-synthetic dimensions, by extending the Potts model to include interactions $J_1$ between atoms adjacent in both real and synthetic space  and studying its finite temperature properties.
For intermediate values of $J_1$, the resulting
phases and phase diagrams are similar
to those of the clock and Villain models, in which three phases emerge. There exists a sheet phase analogous to that found in quantum synthetic dimension models  between the high temperature disordered phase and the low temperature ferromagnetic phase. 
We also employ machine learning to uncover non-trivial features
of the phase diagram using the learning by confusion approach.  The key result there is that the method is able to discern several successive phase transitions.
\end{abstract} 

\maketitle

\section{Introduction} \label{sec: intro}

\vskip0.05in \noindent

\color{black}
Recently experiments on synthetic dimensions 
-- where  transitions between non-spatial degrees of  freedom can be regarded as motion in an additional lattice direction -- have explored physics difficult to realize in real space, such as topological band structures and phases~\cite{lin16,dutt19,kanungo21},  gauge 
potentials\cite{yuan18}, and spatially-resolved non-linear dynamics~\cite{an21}.
Introduced in Ref.~\cite{boada12}, synthetic dimensions were later explored in several platforms, including with synthetic sites formed from momentum states or states of light~\cite{ozawa19}.

We are motivated by synthetic dimensions formed by periodic arrays of ultracold polar molecules or Rydberg atoms in 
optical lattices or microtrap arrays, specifically a two-dimensional (real space) square lattice.   The rotational states of molecules or electronic states of the Rydberg atoms provide the
sites of a synthetic third dimension, and microwaves induce tunneling matrix elements
between the synthetic lattice sites. 
These systems are described by the quantum many-body Hamiltonian~\cite{sundar18,kanungo21},
\begin{align}
\hat H = -&\sum_{i,n} J_n^{\phantom{\dagger}} \big(\, c_{i,n}^{\dagger} c_{i, n+1}^{\phantom{\dagger}}
+ c_{i,n+1}^{\dagger} c_{i, n}^{\phantom{\dagger}} \,\big)
\nonumber \\
+&\sum_{\langle i,j \rangle,n}
V_n^{\phantom{\dagger}}
c_{i,n+1}^{\dagger} c_{i, n}^{\phantom{\dagger}} 
c_{j,n}^{\dagger} c_{j, n+1}^{\phantom{\dagger}}. 
\label{eq:quantumham} 
\end{align}
    The first term describes  tunneling between synthetic sites $n$ at each real-space site $i$, induced by  a  resonant microwaves, while the second term describes an interaction in which atoms or molecules on nearest-neighbor real-space and synthetic sites exchange their synthetic positions.   $\langle i,j\rangle$ indicates a sum over nearest real-space neighbors. Longer-ranged real-space interactions are also present in experiments, but the nearest-neighbor truncation is likely to describe many of the qualitative features of the phase diagram.  

Since  each molecule occupies a unique real-space site, the relevant subspace satisfies
$\sum_{n}  c_{i,n}^{\dagger} c_{i, n}^{\phantom{\dagger}}  = 1$.  In general $J_n$ and $V_n$ depend on $n$, the choice of internal states, and the microwave drives, but in the simplest cases they are independent of $n$~\cite{sundar18}.

Some interesting physics has been predicted for these systems: whereas the hopping $J$ leads to delocalization in the synthetic direction,
the interaction $V$ drives ``sheet formation":  for low temperature and large $V/J$,
molecules on adjacent sites $\langle i,j \rangle$ will share 
a common pair of rotational states $n,n+1$, flattening the system in the synthetic  direction, though with quantum and thermal fluctuations. So far, predicted phase diagrams have been limited to simple variational techniques~\cite{sundar18}, exact solutions in one real dimension~\cite{sundar19}, or certain sign-problem free regions and observables with Quantum Monte Carlo (QMC) methods~\cite{feng22}, which have only begun to be explored.

Equation~\eqref{eq:quantumham} can also be viewed as a large-spin spin models, with the size of the spin $S$ set by the number of synthetic sites $N_s=2S+1$, albeit with interactions and symmetries that are unusual or awkward to describe in the spin language. (especially in the general case where the $J_n$ or $V_n$ have interesting dependences on n) In this way, there is a similarity to the classical Potts or clock models, where there are likewise $q$ possible choices for the variable at each site, with a single one of them `occupied'
in any given configuration. 

However, while corresponding classical spin models would provide important guidance for the physics of the quantum model, the  analogy to classical Potts or clock models has serious shortcomings. The Potts model is invariant under permutations 
 of any of the spin components, a much larger symmetry than the translation symmetry obtained by  translating by one site in the synthetic direction (assuming a large synthetic dimension or periodic boundary conditions). This potentially can cause the Potts model analog  to host  unrealistically large  fluctuations and  varieties of symmetry breaking phases.  The clock model is another analog classical  model, and it enjoys the correct translation symmetry for translating spin indices. However, it has couplings between all spin states, even ones that have large [$O(N_s)$] separations in the synthetic dimension, in contrast to the local nearest-synthetic-neighbor couplings in the synthetic dimension model Eq.~\eqref{eq:quantumham}.  (Potts and clock models are discussed in more detail below.) \cite{wu82,ortiz12,li20}

In this paper, we introduce a classical analog model capturing the salient features of synthetic dimensions that remedies these deficiencies of the Potts and clock models.   We do this by extending the Potts model to include an additional nearest-synthetic-neighbor term, $J_1$, to the Potts model,
as described in the next section, which breaks the permutation symmetry to a translation symmetry (unlike the na\"ive Potts), while preserving locality in the synthetic dimension (unlike the clock model).

Although our motivation is to examine a simple version of a quantum model
of synthetic dimension, as we shall discuss, our variant of the Potts model itself has a rich phenomenology, 
including a non-monotonic behavior of the critical temperature as a function of $J_1$, and the formation of sheet states in analog to the quantum membrane states in the corresponding quantum model.
It also offers an interesting context in which to extend machine learning methods for statistical mechanics.
While our study could certainly be viewed simply as an interesting extension of
a traditional spin model, considering it in the language of synthetic dimension provides
additional context to the work, as well as potential experimental (cold molecule/Rydberg atom)
connections.

 We also note that, interpreted as spin models, synthetic dimensions have obvious parallels in the condensed matter context.  The different
 angular momentum states in $d$ and $f$ electron systems can be considered as a synthetic site index, though in most real solids
one-body effects like crystal field splitting
break the orbital degeneracy and reveal the true ``spin nature" of the model, limiting its interpretation
as an additional dimension. Nevertheless, the classical model we introduce may have commonalities with the interactions in real materials, and suggests phases  that may occur there.

After introducing the model and our primary computational methodology (Sec. \ref{sec: model}), we will examine the physics within a mean field theory (Sec. \ref{sec: MFT}) before presenting numerically exact calculations of properties with Monte Carlo, and inferences for the phase diagram (Secs \ref{sec: EandC}, \ref{sec: sheet}, and \ref{sec: PD}). We will then turn towards a machine learning approach to analyzing the results (Sec. \ref{sec: ML}), and finally summarize our conclusions.

\section{Model and Monte Carlo Methodology} \label{sec: model}

The Potts model \cite{wu82} is a generalization of the Ising model
in which the degrees of freedom  $S_j = 1,2,3, \cdots q$ 
on each lattice site take
on $q$ possible values, with the Ising model
corresponding to $q=2$.  The energy is $-J_0$ when 
adjacent $S_j$ share a common value, and zero otherwise.  A considerable  literature exists
concerning the Potts model, with many generalizations,
including multi-spin interactions, dilution, external fields, 
etc.  The physics of these is well explored, both 
analytically and numerically in Ref.~\cite{wu82}.  Relevant to
the work presented here, the two-dimensional Potts model
exhibits a first order phase transition when $q>4$.

We add a term to the Potts model to describe a situation in 
which the energy is also decreased by $J_1$
if neighboring spins $S_j$ differ by $\pm 1$.  That is,
\begin{align}
E = - J_0 \sum_{\langle ij \rangle} \delta_{S_i,S_j}
- J_1 \sum_{\langle ij \rangle} \delta_{S_i,S_j+1}
- J_1 \sum_{\langle ij \rangle} \delta_{S_i,S_j-1}
\label{eq:extendedpottsmodel}
\end{align} 
We denote by $N=L^2$ the number of spatial sites of 
a lattice with linear dimension $L$.
It's  reasonable to employ a Periodic Boundary Condition (PBC) in both the real dimensions as well as the synthetic dimension.
In this language, the additional $J_1$ term
corresponds to a coupling favoring molecules on neighboring sites
having adjacent positions in the synthetic dimension, 
such as is present in the $V$ term of
Eq.~\eqref{eq:quantumham}.
In the results which follow we set $J_0=1$ as our energy scale.

For $q=3$, Eq.~\ref{eq:extendedpottsmodel}
is the same as the $q=3$ Potts model
with $J_0 \rightarrow J_{\rm eff} \equiv J_0 - J_1$ (for $J_0>J_1)$.  
Thus our model would share all Pott's properties, including a second order transition
at $T_c = J_{\rm eff} / {\rm ln}(1+\sqrt{3})$.
This exact mapping is true only for $q=3$, but is nevertheless suggestive.
We do not study the small $q$ limit further here, since, in
extending the interactions
to include also
$\delta_{S_{i},S_{j}+1}$ and
$\delta_{S_{i},S_{j}-1}$,
for small $q$ the ``range" (three) becomes close to the ``lattice size" $q$
in the synthetic direction.

\begin{figure}[t]
\centering
\includegraphics[width=0.48\textwidth]{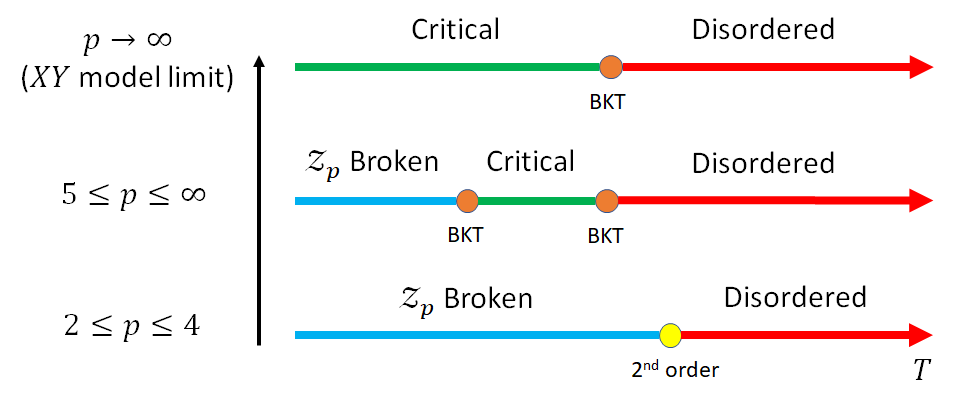}
\caption{ The phase diagram of the $p$ state clock model in two dimensions.
For $p<5$, a single second order transition separates random high $T$ and 
low $T$ phases with true long range order (LRO).  For $p \geq 5$,
this critical point splits into two
distinct BKT transitions, which enclose an extended critical regime 
of power law decay
of correlation functions.  The XY model is recovered
in the $p \rightarrow \infty$ limit where the lower of the two transitions
$T_{c,2} \rightarrow 0$.
Crucial to our work is the presence of an intermediate phase.
\cite{ortiz12}
}
\label{fig:clockPD1}
\end{figure}

The clock model \cite{ortiz12,li20} is another well known description of classical 
phase transitions admitting a controllable discrete set of values 
$n_i=1,2,3,\cdots p$ on each site. It should be noted that, as is the standard notation, the number of Potts degrees of freedom is denoted by $q$, with $p$ being used for the Clock model.
The clock model energy is,
\begin{align}
E= - J_0 \sum_{\langle ij \rangle}
{\rm cos} \bigg( \frac{2 \pi}{p} (n_i - n_j) \bigg) \,\,.
\label{eq:clockmodel}
\end{align}
The clock model shows qualitatively different behavior at different $p$ values, as we will demonstrate happens in
 our model as well. Fig.~\ref{fig:clockPD1} shows the phase diagram of the clock model. Unlike the Ising and Potts
 models, which have single critical points, the clock model, for $p > 4$ exhibits two
distinct transition temperatures, between which there
is a Berezinskii-Kosterlitz-Thouless (BKT) phase. 
The higher-temperature transition occurs at a critical temperature 
$T_{c,1} \sim J_0$, while the lower transition has
$T_{c,2} \sim {\cal O}(1/p^2)$, vanishing in the XY limit $p \rightarrow \infty$. 
These two distinct transitions occur because the $\mathbb{Z}_p$ breaking of the full continuous symmetry of
the XY model is irrelevant for temperatures above $T_{c,2}$ when $p>4$ \cite{jose77,cardy80}.
As we will show in the sections below, Eq.~2 exhibits 
similarity to the clock model in the sense of exhibiting three phases,
with an intermediate `sheet phase' separating the high temperature disordered and low temperature 
$\mathbb{Z}_q$ breaking phases.

To explore the statistical mechanics of Eq.~\ref{eq:extendedpottsmodel}
quantitatively,
we will primarily use the simple classical single site update
Metropolis Monte Carlo method.  We measure observables 
including the internal energy $E$, 
the specific heat $C=dE/dT$, and the average synthetic distance $D$ between neighboring real-space sites, where

\begin{align}
D \equiv \frac{1}{2 N} \sum_{\langle i,j\rangle}  |S_i - S_j|  \,\, .
\label{eq:D_avg}
\end{align} 
The summation runs over all
 neighboring sites $i$ and $j$, and $|S_i-S_j|$ measures the distance of site $i$ and $j$ in
 the `synthetic dimension'.
The normalization is to the number of bonds $2N$  on the square lattice. 
 $D$ is defined to loosely capture a transition into a ``sheet" phase between
 a low temperature phase where all $S_i$ 
are {\it identical} and the high temperature disordered phase. 
We will define this sheet phase more precisely in Section 4.

To characterize the phases further, we define 
\begin{align}
 P_{nn}(d) 
\equiv \frac{1}{2N} \sum_{\langle ij \rangle}
 \delta\big(\, d-|S_i-S_j|\, \big)
\end{align}
which measures the fractional likelihood that the synthetic
 distance $|S_i-S_j|$
between near neighbor sites $\langle i j \rangle$
is exactly $d$.
$D$ and
 $ P_{nn}(d)$ are related by 
$D = \sum_{d} d \, P_{nn}(d)$.

\section{Mean Field Theory} \label{sec: MFT}

The mean field theory of the standard, many-state Potts model 
can be found, for example, in \cite{mftpotts74}.  The addition of a second energy scale for next nearest neighbors presents some new features which we now consider.  For ease of computation, we assign the Potts spin at one site $S$ an accompanying spin variable $\lambda(S)=\omega^S$ ($S=0,1,...,q-1$).  Here, $\omega=e^{2i\pi/q}$ is the complex increment between spin variables.  We then use the identity,
\begin{align}
\delta_{S,S^\prime\pm 1} = \frac{1}{q}\sum^{q}_{k=1} \omega^{k(S-S^\prime\mp 1)} = \frac{1}{q}\sum^{q}_{k=1} \lambda(S)^k \lambda(S^\prime)^{q-k}\omega^{\mp k}
\label{eqn:kronecker}
\end{align}
which allows us to convert Kronecker delta functions in the Hamiltonian to products of spin variables and $\omega$.  Applying this definition to Eq.~\ref{eq:extendedpottsmodel}, the energy between two sites is 
\begin{align}
E_{S,S^\prime}=
-\frac{1}{q}\sum_{k=1}^{q}
\big[J_0+J_1\big(\omega^k+\omega^{-k}\big)\big]\lambda(S)^k\lambda(S^\prime)^{q-k}
\end{align}
We introduce the mean field by defining a set of $q-1$ order parameters,
$\langle \lambda^k\rangle=R\,e^{ik\theta}$ \cite{mftpotts74}.  $\theta$ serves as the `alignment' of the mean field and can take on values of $\theta=0,\frac{2\pi(1)}{q},...,\frac{2\pi(q-1)}{q}$; if $\theta$ is $\frac{2\pi(4)}{q}$, then the field is aligned relative to $S=5$ (akin to an origin).  As the Potts model views neighboring sites as being the same spin, off by one, or of any other value, we are free to choose our field alignment and hence set $\theta=0$ for convenience.  Replacing the spin variable $\lambda(S^\prime)^{q-k}$ with order parameter $\langle \lambda^{q-k}\rangle=R$ leaves us with
\begin{align}
& H_S=
-\frac{nR}{q}\sum_{k=1}^{q}
\big[J_0\omega^{Sk}+J_1\big(\omega^{(S+1)k}+\omega^{(S-1)k}\big)\big] \\
& E_{S} = - nR
\big[J_0\delta_{S,0}
+J_1\big(\delta_{S-1,0}+\delta_{S+1,0}\big)\big]
\label{eqn:newhamilt}
\end{align}
where $n$ is the number of nearest neighbor sites ($n=4$ for a square lattice).  Taking the expectation value $\langle\lambda\rangle$ 
and setting it equal to $R$, we arrive at the self-consistent equation
\begin{align}
R=\frac{\text{Tr}\big(\lambda e^{-\beta H_S}\big)}{\text{Tr}\big(e^{-\beta H_S}\big)}=
\frac{\big(e^{\beta nJ_0R}-1\big)+2\big(e^{\beta nJ_1R}-1\big)
\cos\frac{2\pi}{q}}{e^{\beta nJ_0R}+2e^{\beta nJ_1R}+q-3}
\end{align}
which at $J_1=0$ reduces to the self-consistent equation of the standard Potts mean field theory \cite{mftpotts74}.  

\begin{figure}[t!]
\centering
\includegraphics[width=0.50\textwidth]{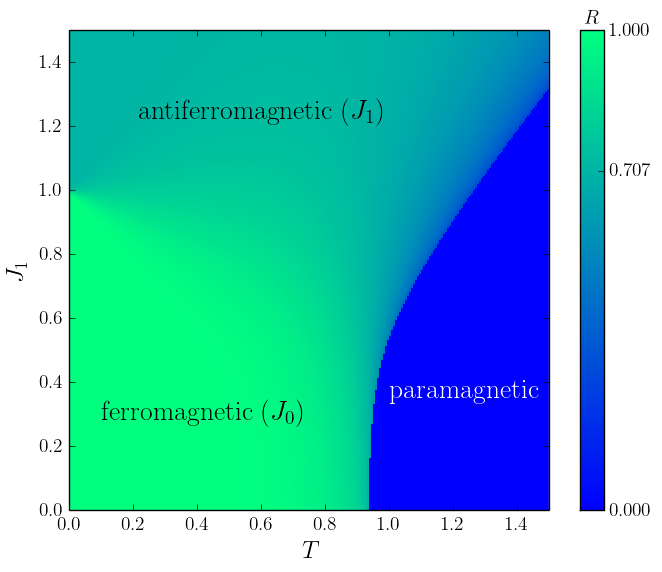}
\caption{
Heat map of order parameter $R$, showing the mean field phase diagram of the generalized Potts model for $q=8$.  The value of $R$ serves as a measure of how aligned a spin is with the field via $R={\rm Re}( e^{2 \pi i\Delta S/q})$, so a system with identical spins (ferromagnetic) has $R=1$.  Likewise, if the spin differs from the field by a value of $\pm1$ (antiferromagnetic) for $q=8$, then $R=0.707$ as seen in the diagram.  In this and all subsequent figures, energies and temperatures are in units of $J_0$.}

\label{fig:MFTPD}
\end{figure}

The resulting phase diagram with $q=8$ is given in the heat map of Fig.~\ref{fig:MFTPD}. The blue region shows the high temperature, disordered phase where the order parameter $R=0$.
For $J_1/J_0 < 1$ there is the `conventional' low $T$ ferromagnetic phase of the Potts model (bright green) 
where Potts variables on different sites are 
identical.
However a second ordered phase appears (cool green) for $J_1/J_0>1$ in which $R={\rm Re}( e^{2 \pi i/q})
= \sqrt{2}/2=0.707$ for $q=8$.  In this region, the Potts variables on adjacent sites differ
by $\pm 1$.  We refer to this as an `antiferromagnetic' regime since adjacent spins are forced to take distinct values.
We will argue, based on the approximation-free Monte Carlo solution in the next section, in which
we can take more sensitive `snapshots' of the spin configuration, that an additional intermediate
regime arises in which the Potts variables on adjacent sites are either identical, or differ by $\pm 1$. In the heat map of Fig.~\ref{fig:MFTPD}, this behavior is seen in the dark blue region that emerges for $J_1/J_0=1$, where $0.707<R<1$ as if the system was allowing for the coexistence of nearest- and next nearest-neighboring interactions between ferromagnetic and antiferromagnetic ordering.
The critical point at $J_1/J_0=0$ is consistent with the analytic value
$T_c = 12/(7 \, {\rm ln}7) = 0.8810$ for $q=8$.

\begin{figure}[t]
\centering
\hspace*{-0.10in}
\includegraphics[width=0.48\textwidth]{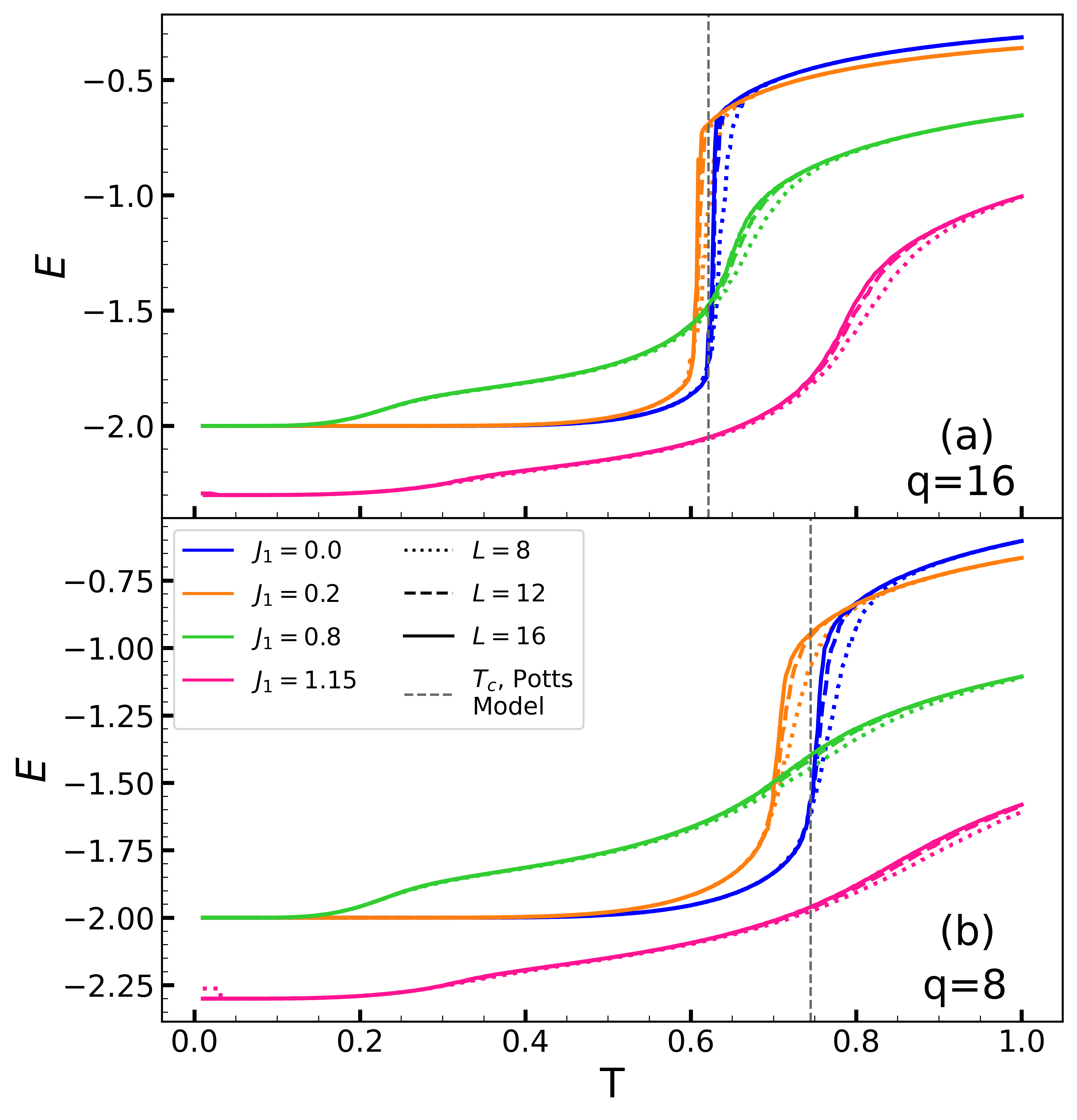}
\caption{
Internal energy per site $E(T)/N$ for $q=16$ 
(panel a) and $q=8$ (panel b) for different $J_1$ (from bottom to top at $T=1.0$, $J_1=1.15,0.8,0.2,0.0$ respectively)
and lattice sizes $N=8 \times 8$, $12 \times 12$, 
and $16 \times 16$. Finite size effects are minor, 
and $E(T)/N$ appears converged on the largest lattice. 
} \label{fig:energy1}
\end{figure}

\section{$E$ and $C$ for $q=8$ and $q=16$} \label{sec: EandC}

We turn now present results obtained with Monte Carlo.

It is known for
the pure clock model that $q=5$ is a critical value for the
appearance of a BKT phase at intermediate temperature, 
and that the phase diagram is qualitatively the same
 for any $q>5$, i.e.~differing only in the range of temperature
over which the intermediate phase is stable, as seen in 
Fig.~\ref{fig:clockPD1}.
Similarly, we observe a intermediate regime that differs from the ground state behavior for $q>6$ in our model as well. 
To explore the novel intermediate region, we focus our study on just two 
values: $q=8$ and $q=16$.  
This choice also allows us to gain some insight into quantitative changes with $q$ without excessive redundancy.
We will demonstrate later that instead of creating pairs of vortexes as in the clock model, a `sheet' is formed in the intermediate temperature region when the $J_1$ interaction sufficiently strong.

We begin by showing the energy per site $E(T)/N$ for 
$J_1=0.0, 0.2, 0.8, 1.15$ with $q=16$ 
(Fig.~\ref{fig:energy1}a)
and $q=8$ 
(Fig.~\ref{fig:energy1}b).  
For $J_1<J_0=1$ the ground state has all the sites taking 
the same spin value, and an energy per site $E_0/N  = -2J_0$. We will call this ground state the Ferromagnetic (FM) phase. 
For $J_1>J_0=1$,
it is energetically favored to have adjacent sites with $S_j = S_i \pm 1$
so that $E_0/N  = -2 J_1$ at $T=0$. 
We call this the Antiferromagnetic (AFM) phase. 
In both panels, 
the known first order transition at
$T_c=J_0/{\rm ln}(1 + \sqrt{q})$
of the conventional Potts model
($J_1=0.0$) is shown for comparison.
This first order character remains evident at small $J_1$, however, occurring at a (slightly) reduced $T_c$.
We interpret the lowering of $T_c$ to arise from
the fact that $J_1$ initially competes with the
tendency of $J_0$ to make all values of
$S_j$ identical.

\begin{figure}[t]
\centering
 \includegraphics[width=0.48\textwidth]{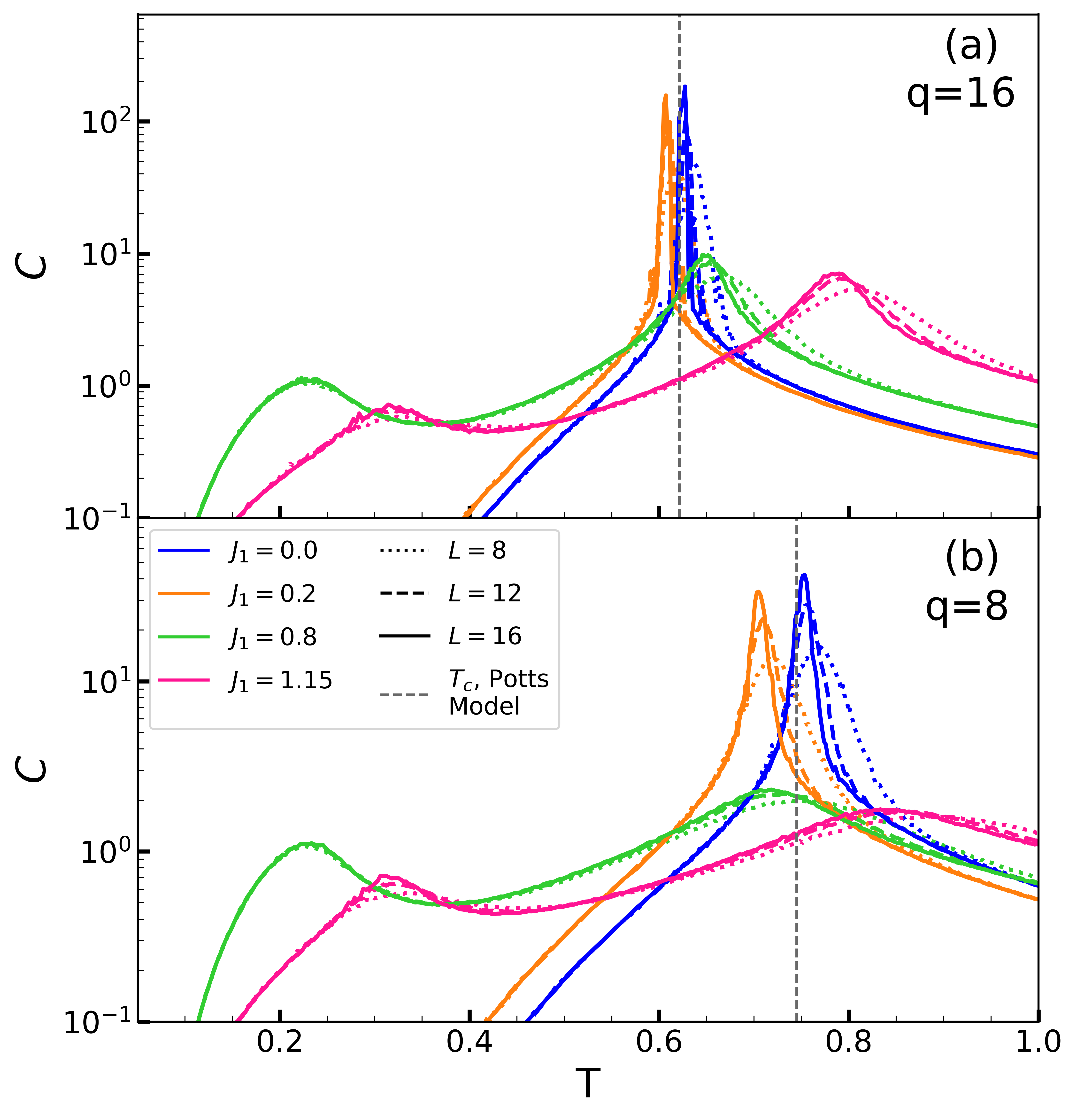}
\caption{
Specific heat $C(T)$ (on a log scale) for $q=8$ and $16$ with different 
choices of $J_1$ (the curves whose intersection with the $T$ axis from left to right represent $J_1=0.8,1.15,0.2,0.0$ respectively) and lattice sizes $N=8 \times 8$, $12 \times 12$, 
and $16 \times 16$. The dashed gray lines denote analytic values
for $T_c$ for the Potts model ($J_1=0$). The behaviors of $C(T)$ 
for $q=16$ (panel a) and $q=8$ (panel b) are 
qualitatively the same. The size of 
heat capacity peak is lower by over an order of magnitude 
for $J_1=0.8$ compared to $J_1=0.0, 0.2$, a signature of the change of order of 
transition from first to second.
}
\label{fig:specificheat1}
\end{figure}

As $J_1$ increase further, however, $T_c$ begins
to grow. Here our picture is that the two interactions behave in a more cooperative
way, raising the critical temperature. Notably, it can be seen that when $J_1$ is small, $T_c$ predicted by mean field theory is slightly higher than that predicted by Monte Carlo data, which is to be expected, as mean field theory overestimates $T_c$ by neglecting fluctuations. Simultaneously, the evolution of $E(T)$ becomes more gradual, suggesting a continuous phase transition, or even a crossover.
Studies on a finer mesh of $J_1$ indicate
the change of order of transition appears to coincide with the value of $J_1$ for which $T_c$ is minimized. 
As can be seen from the $J_1=0.8$ curve, 
aside from the now-continuous transition at $T \sim 0.7$, a second 
feature emerges at $T \sim 0.2$.
This is a first indication of the existence of an intermediate regime.

The change from first-order character is further revealed in the specific heat.
Figure \ref{fig:specificheat1} gives $C(T)$ for 
the same sequence of $J_1$, 
again for $q=16$ and $q=8$.
For $J_1=0.0, 0.2$, the first-order discontinuity of 
$E(T)$ in the thermodynamic limit
is signalled by the very large value of the finite lattice derivative,
exceeding by more than an order of magnitude 
the peaks at  $J_1=0.8, 1.15$ where the
transition is second order. 
Framed more precisely, at a first order transition 
the specific heat on a finite lattice scales as the
volume of the system, $C(T) \sim N=L^2$,
whereas for a second order transition $C(T) \sim L^{\alpha/\nu}$.
Here $\alpha$ and $\nu$ are the critical exponents for the specific heat and correlation length respectively.
As a consequence, the specific heat peaks are much smaller in the second order region.

As noted already in the data for $E(T)$,
Fig.~\ref{fig:energy1},
for $J_1=0.8, 1.15$ an additional, lower temperature, 
specific heat peak emerges.  This suggests the existence
of an intermediate regime between the ordered phase 
at low temperature and the disordered
phase at high temperature. We shall focus our discussion to 
the intermediate regime in the next section. 
Given the qualitative similarity of
the results for $q=8$ and $q=16$, 
in the next section we will restrict our analysis to the larger value of $q=16$.

\section{Sheet Formation} \label{sec: sheet}

\begin{figure}[t!]
\centering
 \includegraphics[width=0.48\textwidth]{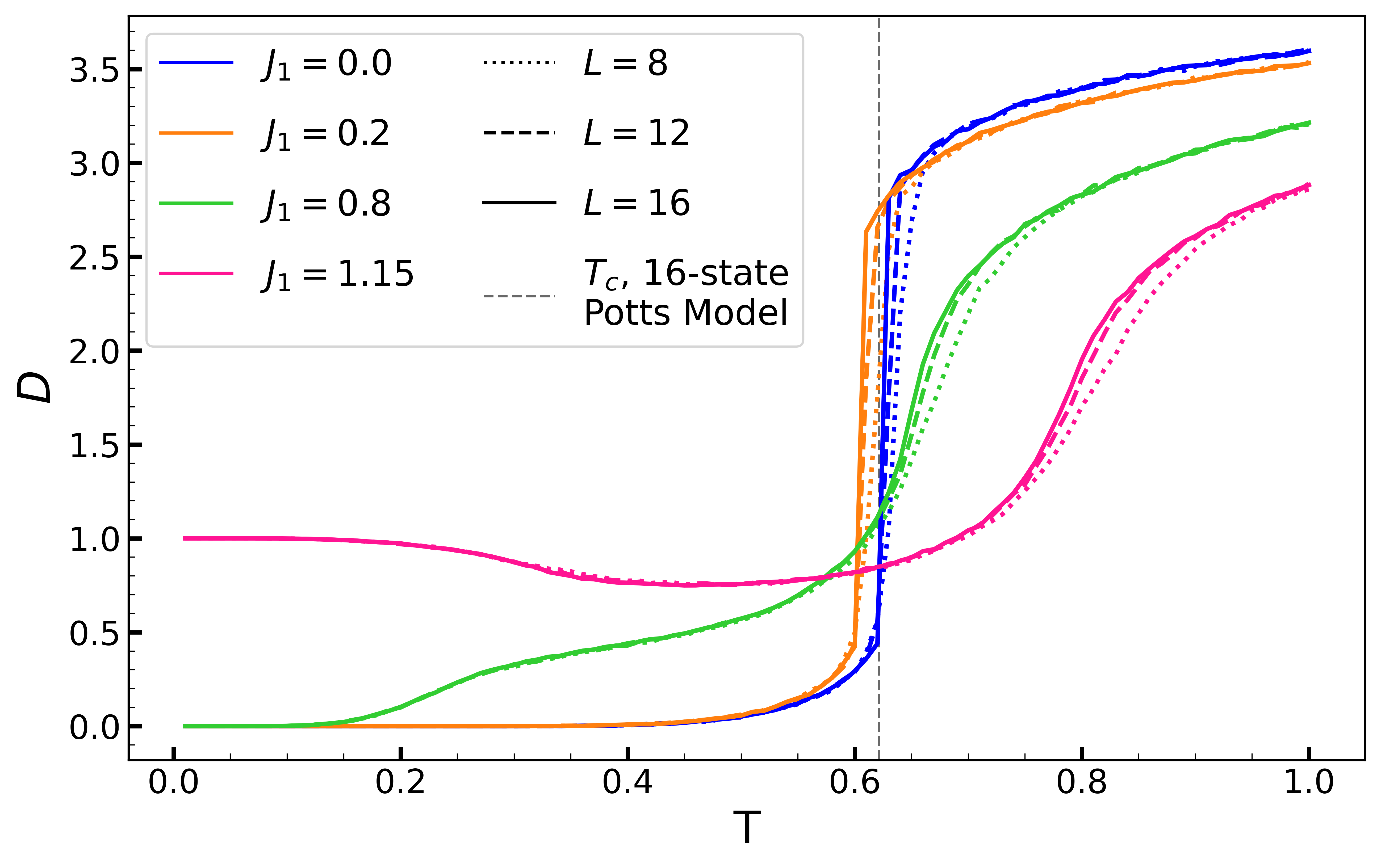}
\caption{
Temperature dependence of 
$D$ for different $J_1$ (the curves 
from bottom to top at $T=1.0$ represent $J_1=1.15,0.8,0.2,0.0$ respectively) and lattice sizes of $8\times8$, $12\times12$ and $16\times16$. For 
$J_1=0.0$ and $0.2$, the first order signature is evident. 
For $J_1=0.8$ and $1.15$, there exists an intermediate 
plateau that is evident for a sheet phase between the 
ground state and the high temperature disordered phase. 
}
\label{fig:nndistoverQ}
\end{figure}

We now turn to examining the nature of the intermediate region, 
in which we believe a `sheet' forms,
analogous to that of the quantum model \cite{feng22}.  
Fig.~\ref{fig:nndistoverQ} shows the
 average distance $D$ in synthetic dimension between nearest neighbors 
on the lattice, as a function of $T$, for $J_1=0.0,0.2,0.8$ and 
$1.15$, that is, horizontal sweeps 
of the phase diagram of Fig.~\ref{fig:MFTPD}.
For all $J_1$ values, we see $D$ asymptote at high temperature to 
$D=q/4$, which indicates
a disordered phase where the spins randomly (probability $1/q$) and independently occupy different synthetic position.. 
(Note our use of PBC in the synthetic direction imposes
a maximal value $D=q/2$.)

For $J_1=0.0$ and $0.2$,  $D$ increases from $0$ to the high 
temperature limit of $q/4$  through a direct jump, 
suggesting a single first order phase transition from the 
ordered phase to the disordered phase, like the Potts model. 
However, for $J_1=0.8$, $D$ shows a plateau which coincides
with the temperature range between the two 
specific heat peaks in Fig.~\ref{fig:specificheat1}. In this 
region, $J_1$ is sufficiently strong to 
allow neighboring spins $S_i$ and $S_j$ to take on values 
$S_j = S_i \pm 1$
in addition to $S_j=S_i$ favored by $J_0$. 
The precise value of $D$ depends on $J_1$, which will
control the relative likelihood of
$S_j = S_i \pm 1$ and $S_j=S_i$.   If those
cases are all equally probable, for example, then
$D \sim 2/3$.  
We will refer to this as the sheet phase, because spatially 
neighboring spins exist in a range of three adjacent values, and, 
as we shall see, this persists to larger separations, so
that the configuration 
looks like a sheet in $2+1$ dimensions. This is to be distinguished from the Ferromagnetic phase, where all spins are identical.

For $J_1=1.15$, the ground state is different. 
$D=1$ at low temperature is expected now since all the 
neighboring spins favor taking values that differ by one.
This change of ground state can be 
checked in the $T=0$ limit of $E$ in Fig.~\ref{fig:energy1}, 
which takes on the value of $E=-2J_1$ as expected. 
With increasing $T$, 
neighboring spins in the system start to take on values that are the same, thus the system enters the sheet regime, before 
going into the disordered phase at high temperature, as was the case with $J_1=0.8$. That the intermediate $T$ region 
of $J_1=0.8$ and $J_1=1.15$ are of the same nature, and no 
transition is between them, will be corroborated below.

The existence of the intermediate sheet region can be qualitatively understood by looking at the system's free energy.
For $J_1<1$, at very low but finite temperature, excited states become accessible. The lowest-energy excitations are the ones where one site takes a spin value that differs by one from its neighbors'. This costs an energy of $\Delta E = (J_0-J_1) $. Comparing to a Potts model with $J=(J_0-J_1)$, it's evident that our model should retain its FM order for a range of finite low temperatures, since all of its excitations cost more energy than that of the Potts model's, which is know to have a FM ground state at low temperature. However, upon increasing $T$, a manifold of excited states that we call 'sheet state' will proliferate. The sheet state refers to a state where all the neighboring spins take on spin values that are either the same or differ by one. Such states have extensive energy $\Delta E \propto (J_0-J_1) $, and entropy per site $\Delta S = \ln 3$. The latter can be understood by a counting similar to that of degenerate ground states for $J_1 > 1$, which is outlined in the Appendix  \ref{Appendix}. This means sheet states will have a free energy contribution $\Delta F \propto (J_0-J_1) -T \ln 3$, which goes negative at $T \gtrsim (J_0-J_1) / \ln 3$. We will show in the next paragraph that the system makes a transition into a sheet-state regime at a finite temperature for $J_1 \gtrsim J_{1c} \sim 0.4$ for $q=16$. 
As we continue to increase $T$, the system will go into a disordered state as expected.
We recall that it was also observed in $E(T)$ and $C(T)$ plots that at the same $J_{1c}$ where sheet phase appears, the transition to the high temperature disordered phase becomes continuous. We argue that this is plausible as the system in the sheet regime could have different symmetry from its ground state. This also raises the question of whether the transition from the ground state to the sheet regime is a true phase transition. We will come back to this question when discussing the phase diagram.

To support the existence of the sheet regime further, 
we measure $P_{nn}(|S_i-S_j|)$.
In the upper panel, we see $P_{nn}(0) \sim 1$ 
while both $P_{nn}(1)$ and 
$P_{nn}(2) \sim 0$ at low $T$ for $J_1=0.8$. 
This tells us all the spins are taking the same value as they 
should in the FM phase. As we increase $T$ past 
the temperature of the lower heat capacity peak, 
$P_{nn}(0)$ starts to drop significantly, 
while $P_{nn}(1)$ increases.
Meanwhile $P_{nn}(2)$ is still very close to $0$. 
This is to say, in this region, neighboring spins are equally 
likely to take values that are identical and off by one, but 
they dislike taking values that are further separated,
the defining characteristic of the sheet phase, and consistent with
Fig.~\ref{fig:nndistoverQ}.  As we 
increase $T$ further, all the curves go to the same limit, 
resembling the disordered phase where all spin values are equally likely. 

Similarly, in the upper panel, the low $T$ limit of $J_1=1.15$ 
is showing a clear feature of the AFM phase. 
In the intermediate $T$ region between the two heat 
capacity peaks, the decrease of $P_{nn}(1)$ 
and the increase of $P_{nn}(0)$ 
are both significant, while $P_{nn}(2)$ 
is kept at a low level. This  
is similar to the behavior at $J_1=0.8$. 
The difference 
is that here we're coming to the sheet regime from an AFM phase 
instead of a FM phase.  
This picture will be validated by the machine learning 
method explained in the next section.

\begin{figure}[ht]
\includegraphics[height=5in,width=3.3in]{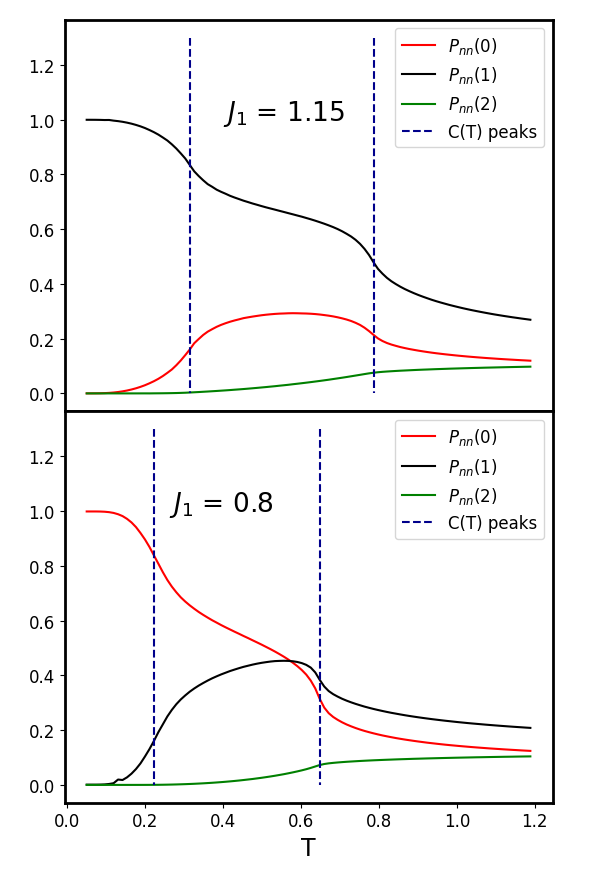}
\caption{
$P_{nn}(0)$, $P_{nn}(1)$, 
and $P_{nn}(2)$ vs temperature for 
$J_1 = 1.15$ and $J_1 = 0.8$. 
For both values of $J_1$, $P_{nn}(2)$ is the bottom curve, indicating that adjacent sites rarely have synthetic distance of 2. Furthermore, both plots show rapid changes near the specific heat peaks. Furthermore, both plots show an intermediate phase in which, for neighboring sites, the synthetic distance is allowed to be either 0 or 1, therefore supporting the existence of the sheet phase.
}
\label{fig:Pij}
\end{figure}

\section{Machine Learning Approaches} \label{sec: ML}

 Both supervised and unsupervised machine learning algorithms have been used recently to 
 find critical points of classical and quantum models of magnetism
 \cite{wang16,hu17,carrasquilla17,zhang19,canabarro19, nieuwenburg2017, lee2019}.
Many of the most impressive applications of machine learning are those
using unsupervised algorithms, since no prior knowledge of the phase diagram
is needed. Recently, for example, phase transitions of the 
$q$-state Potts model were analyzed with a series of 
unsupervised techniques including Principal Component Analysis 
and k-means clustering, among others \cite{Tirelli2022}.  
While the results of these methods were certainly impressive, 
we now explore the use of an unsupervised method
 known as the learning by confusion (LBC) approach, which uses neural networks 
instead of dimensional reduction or clustering algorithms. 
This method is becoming very widespread, and was even used 
to examine the $q$-state Potts model in a recent study \cite{Laskowska2022}.
In this section, we outline the LBC method and the details of our implementation, 
and subsequently present our results.

The LBC algorithm, while being completely unsupervised, can be thought
of as a series of attempts at supervised machine learning. Raw spin
configurations generated from Monte Carlo are used as input data, which
are then purposely given incorrect labels. In particular, `fake'
critical points $T_c^\prime$ are postulated:  for each, 
all configurations corresponding
to temperatures below $T_c^\prime$ are given a label of 0, and those above
are given a label of 1. The data are then split into training and
testing data sets. A neural network is trained with the labeled
data, and the accuracy is computed against the test data split.
As each new $T_c^\prime$ is chosen, the neural network is 
re-trained, and a new accuracy
is then calculated. 
 
The key observation  is the following:  When $T_c^\prime$ is very large
(i.e.~higher than all the temperatures sampled in the Monte Carlo), 
all data, training and test, are labeled 0, and the neural net can learn 
that trivial labeling with 100\% accuracy.
The same is true for 
$T_c^\prime$ 
very small.
Likewise, when $T_c^\prime$ 
arrives at 
a value close to the true transition temperature,
the neural network will be able to classify most of the
configurations correctly, yielding a very high accuracy. 
For other values of
$T_c^\prime$, 
the neural network will fail to distinguish some configurations 
from each other, resulting in comparatively low accuracies. It follows,
then, that the accuracy vs temperature plots will have a `W' shape, with
the middle peak corresponding to the correct $T_c$. In cases with two phase transitions, accuracy vs temperature plots result in a double `W' shape, with the two peaks corresponding to the two critical temperatures.
This process can also be done holding temperature constant, 
and varying other degrees of freedom such as $J_1$.

Although the LBC method has had major success in recent years, it is important to note that, because we are using finite sized lattices, it is possible that the accuracy peak corresponds to a crossover rather than a true phase transition. While there is work indicating that the shape of the peak corresponds to the order of the phase transition \cite{Laskowska2022}, it is still certainly possible that a peak corresponds to a crossover. As such, we turn to more traditional methods to more closely characterize the nature of the LBC accuracy peaks.

\begin {figure}[t]
\includegraphics[height=2.5in,width=3.0in]{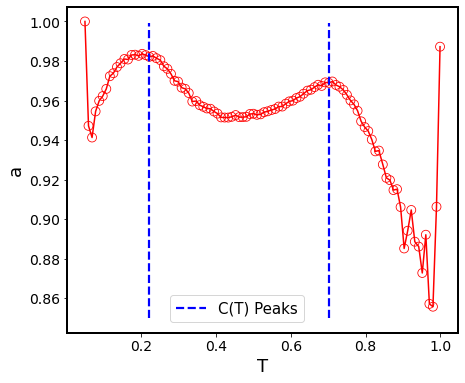}
\caption{Output of the LBC method for $q=8$, 
$J_1=0.8$. Each point on the plot corresponds
to a different CNN for which the accuracy $a$ 
was evaluated after training. When near the true
transition temperature, the CNN is better able to 
classify the data into two groups, resulting
in higher accuracy. Peaks were found by removing the left and right most points and fitting a double Gaussian to the remaining data. For these parameters, the LBC method was able to find agreement
with traditional methods to within a 6\% difference.}
\label{fig:cnnvsfc}
\end{figure}

Traditionally, the LBC method is used with a fully connected neural
network, which is expressive enough to distinguish phase transitions
in many cases \cite{nieuwenburg2017, lee2019}. In our case, however,
such neural networks performed quite poorly, so convolutional neural
networks (CNNs) were used instead, which are much better at picking
out subtle spatial information from the data. It may be possible to
get successful results with fully connected neural networks, but more
data or a more complex training protocol may need to be used, slowing
down the algorithm dramatically. Fig.~\ref{fig:cnnvsfc} 
shows an example of the LBC method using CNNs, 
illustrating congruent results with traditional methods.

\begin {figure*}[t]
\includegraphics[height=4in,width=6.5in]{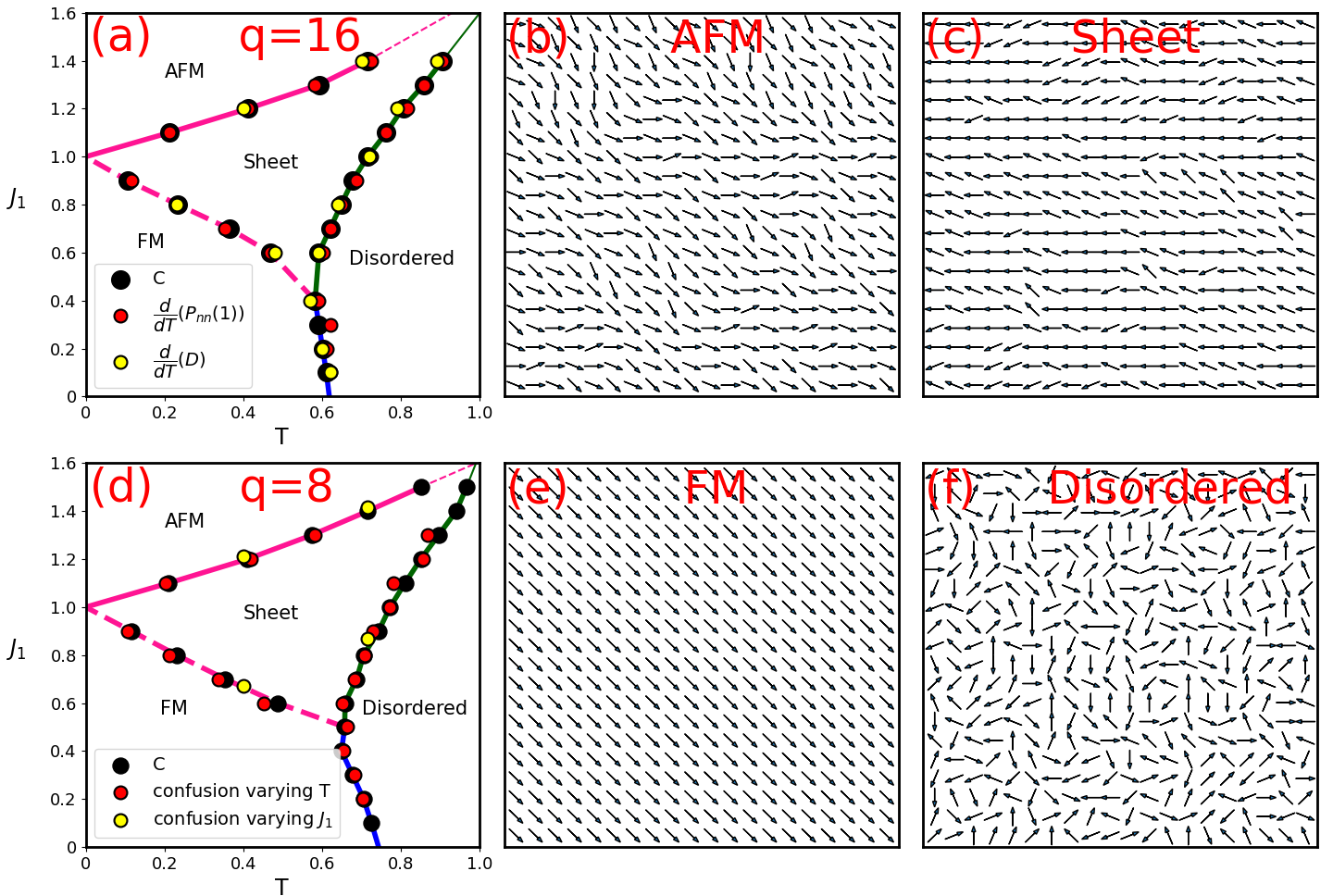}
\caption{Panel (a) shows the phase diagram for our model for $q=16$, featuring a comparison of specific heat, $P_{nn}(1)$, and $D$. Since the phase boundaries correspond to the temperatures in which $P_{nn}(1)$ and $D$ change most rapidly, points associated with these quantities were found by locating the extrema of the derivatives $\frac{d}{dT}\left(P_{nn}(1)\right)$ and $\frac{d}{dT}\left(D\right)$.
For small $J_1$ there is a single first order transition to a 
ferromagnetic phase (all spin variables identical). For intermediate 
$ 0.6 \lesssim J_1 \lesssim 1.7$  two distinct transitions occur. The low temperature boundary is likely a crossover for $J_1 < 1$ (dashed magenta boundary) and a continuous phase transition for $J_1 > 1$ (solid magenta boundary). The higher temperature boundary is likely a continuous phase transition as well (green boundary). 
The intermediate regime is characterized as a `sheet' region where the 
neighboring sites have spin variables differing by $\pm 1$. 
Panel (d) shows a similar phase diagram for $q=8$, comparing specific heat peaks to LBC peaks. 
The remaining panels show snapshots of the four states in the phase diagram, all for $q=16$.
Panel (b) shows a snapshot of the anti-ferromagnetic phase in which near neighbor sites are required to exhibit $S_j = S_i \pm 1$, and (c) shows that of the sheet regime. Panel (e) shows a snapshot of the ferromagnetic phase, and (f) displays that of the disordered phase.
}
\label{fig:PD}
\end{figure*}

We tested the LBC approach on data generated from Monte Carlo simulations
using $q=8$, as it is sufficiently high for the existence of the sheet regime, but moving to larger $q$ values would increase the time for computation, 
as explained in the Appendix. We also tested the LBC scheme at many different values
of $J_1$ to get more comprehensive results. As expected, LBC peaks and specific heat peaks were in strong agreement, both in cases with and without the sheet regime. Fig.~\ref{fig:PD}
shows the resulting phase diagram generated from many 
runs similar to that of Fig.~\ref{fig:cnnvsfc}. 
In addition to sweeps across temperature, we also 
present multiple runs holding temperature constant, 
and sweeping across $J_1$ to further validate our results.

The use of the LBC approach applied to this model is not to claim superiority over traditional methods such as C(T) peaks, but rather to highlight its application to a specific problem in which the phase diagram was unknown. By benchmarking the LBC method in this way, we aim to show its effectiveness and reliability. The ultimate goal is to help expand the toolkit available for dealing with novel or poorly understood systems. In this way, we hope to underscore the potential of the LBC approach to provide new insights, which will become useful in areas where more traditional methods might face challenges.
 
\section{Phase diagram} \label{sec: PD}
Based on the analysis in the above sections, we present 
in Fig.~\ref{fig:PD} the phase diagrams of our model 
for $q=8$ and $q=16$, as well as snapshots of states 
in each of the phases. As we found in Sec. \ref{sec: EandC}, $q=8$ 
and $q=16$ have qualitatively the same behavior, 
which is confirmed in panels (a) and (d). While there 
are slight differences in the specific transition temperatures, 
both phase diagrams have very similar structures. 
Lower $q$ values such as $q=4$, however, are likely 
to yield much different phase diagrams, as is the case 
in the $q$ state clock model \cite{ortiz12}.

At sufficiently low $T$, the system stays at its FM or 
AFM ground states for $J_1<1$ and $J_1>1$ respectively, 
snapshots of which are also shown in Fig.~\ref{fig:PD}. 
In the FM phase, all spins line up, whereas in the AFM phase, 
spins deviates by exactly one spin value from that 
of its neighbor's, rendering a fluctuating-looking snapshot. 

Below a critical value of $J_{1c}$ ($J_{1c} \sim 0.4$ 
for $q=16$ and $J_{1c} \sim 0.5$ for $q=8$), the system 
acts like a Potts model, which transitions into the 
disordered phase through a first order phase transition, 
as indicated by the blue line in the phase diagram. 
Increasing $J_1$ only slightly lowers $T_c$, which can be 
understood since $J_1$ makes it harder for the system to 
develop ferromagnetic order. 

Above the critical value of $J_{1c}$, as we increase $T$, 
the system develops a sheet regime,
before a phase transition into the disordered phase. A snapshot of the sheet 
regime is shown, where neighboring spins are able to 
take the same value as well as off-by-one values. 
For this reason, if we treat the spin values as a 
synthetic dimension, we'll see a sheet in the 2 spatial + 
1 synthetic dimension. For $J_1<1$, the smooth transition in
$E(T)$ in Fig.~\ref{fig:energy1} and the size independence of the peak in $C(T)$ in Fig.~\ref{fig:specificheat1} indicates that the boundary between the FM and sheet region is either a crossover or a KT transition. Similar size independence of heat capacity is observed in the $XY$ model as well \cite{nguyen2021superfluid}. This is marked as 
a dashed magenta boundary in the phase diagrams.
For $J_1>1$, the transition from the AFM to the sheet 
region looks continuous as well if we look at the $E(T)$ plot in Fig.~\ref{fig:energy1}. However, we believe it's likely a second order phase transition as Fig.~\ref{fig:specificheat1} shows clear size dependence of the heat capacity peak for $J_1=1.15$, as opposed to $J_1=0.8$. Furthermore, the FM and AFM phases are distinct, in that there should be a phase transition dividing the two regions. This provides further evidence that the boundary from the AFM phase to the sheet regime is a phase transition as opposed to a crossover. We have marked the phase boundary pink to differentiate 
from the first order transition happening between the FM and sheet phase. 

\begin {figure}[t]
\includegraphics[width=0.48\textwidth]{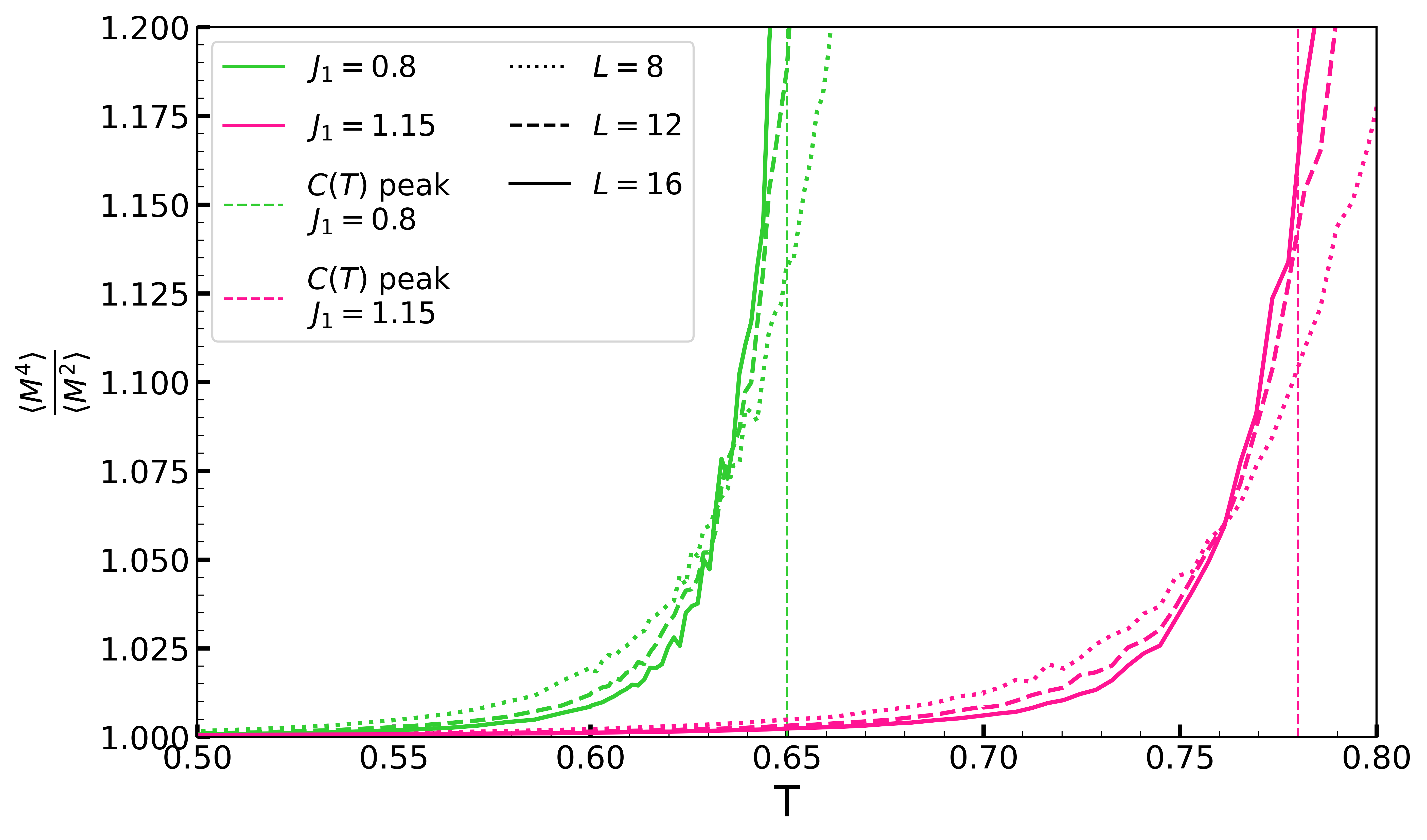}
\caption{Binder ratio $\frac{\langle M^4 \rangle}{\langle M^2 \rangle^2}$ for $L=8,12,16$. Crossing points are seen for $J_1=0.8$ and $1.15$. Heat capacity peak positions for the same $J_1$ values are plotted as vertical dashed lines for comparison, in good agreement with the crossing point of Binder ratio. A slight deviation is expected as consequence of finite size effects for heat capacity peaks.}
\label{fig:Binder}
\end{figure}

Finally for all $J_1>J_{1c}$, the system will 
eventually transition to a disordered phase through 
a second order phase transition, indicated by the 
green boundary in the phase diagram. We are be able to corroborate this claim while locating the transition temperature more accurately by using the Binder ratio $\frac{\langle M^4 \rangle}{\langle M^2 \rangle^2}$. Here $M$ is defined as $M=\frac{1}{N} \sum_j e^{\frac{2 \pi S_j i}{q}}$. It can be seen from Fig.~\ref{fig:Binder} that the positions of the corresponding heat capacity peaks are very close to the critical temperatures indicated by the Binder ratio crossing points. However we should mention that the Binder ratio fails to capture other phase boundaries, thus we're not showing the Binder ratio in a broader temperature range here. 

\section{Conclusions} \label{sec: conc}

In this paper we have explored the phase diagram of a variant
of the Potts model in which the energy is lowered not only when the degrees of freedom 
$S_i = 1, 2, \cdots q$
on adjacent spatial sites are identical,
but where there is an additional energy lowering when they differ
by $\pm 1$, that is when $S_j = S_i \pm 1$.
Such a situation mimics that of experiments with ultracold dipolar particles in synthetic dimensions, for example ultracold polar molecules in an optical
trap in which each molecule has a ladder of possible rotational states
with $z$ component of the angular momentum $m_i$ of molecule $i$, 
and inter-molecular dipole interactions favor $m_j = m_i \pm 1$.

Although a precise connection between the two models is limited
by the absence of quantum fluctuations in our classical
analog, our results reveal several interesting qualitative links.
In particular, we observe the emergence of a low temperature sheet regime which is very similar to one of the defining characteristics of the molecular system. 

Prior studies \cite{glumac98} of the three state 
Potts model have considered the effect of the range of the interaction
$V(r) \sim 1/r^{1 + \sigma}$ on the order of the transition, finding
first order character for $\sigma \lesssim 0.7$.  The context was that of
a lattice with one (physical) dimension and long range (power law decaying) interactions.
Our specific perspective here has been to formulate a classical model which
serves as an analog of quantum models of sheet formation in synthetic dimension.
Our work, which incorporates
an energy which is lowered 
when adjacent sites (in two `physical' dimensions) have variables which
are identical (the usual Potts model) or differ by $\pm 1$ (our extension)
can be interpreted as a similar
extension of  the `range' of the interaction, but only
in the third `synthetic' direction.

Finally, we emphasize that
determination of the phase diagrams of such models poses an interesting new
challenge to machine learning and related information-theoretic \cite{negrete18,negrete21}
approaches being developed for statistical mechanics.
We have shown here that learning by confusion \cite{nieuwenburg2017, lee2019}
provides a powerful and accurate
means to gain insight, and, significantly,  that it is able to
capture a `double-$W$' structure of the accuracy which might be expected
in the presence of two successive finite temperature transitions.

\section{Acknowledgements}

M.C., Y.Z., M.C, and R.T.S
acknowledge support from the U.S.~Department of Energy, Office of Science,
Office of Basic Energy Sciences, under Award Number DE-SC0022311. K.H. acknowledges support from the Welch Foundation (C-1872) and the National Science Foundation (PHY-1848304). K.H.
thanks the Aspen Center for Physics, supported by the
National Science Foundation grant PHY-1066293, and
the KITP, supported in part by the National Science
Foundation under Grant PHY-1748958, for their hospitality while part of this work was performed.

\bibliography{main03}

\clearpage
\renewcommand{\theequation}{S\arabic{equation}}
\setcounter{equation}{0}
\begin{center}
{\large \bf Appendix:
Classical Analog
of Quantum Models in Synthetic Dimension}
\vspace{0.3cm}
\end{center}
\vspace{0.6cm}

\beginsupplement \label{Appendix}

In this Appendix we consider details 
which provide additional insight into the model of Eq.~\ref{eq:extendedpottsmodel}
including 
(1) an evaluation of the ground state entropy, which we show
to be extensive; and
(2) details of the machine learning approach.

\vskip0.10in \noindent
{\bf A. Macroscopic Ground State Entropy}
\vskip0.07in \noindent

When $J_0$ is dominant, there is a $q$-fold degenerate ground state
consisting of all sites sharing a common spin value.  However, this degeneracy is
`small' in the sense that the entropy per site $s(T=0) = \frac{{\rm  ln} \, q}{N}$
vanishes in the thermodynamic limit (i.e.~fixed $q$ and large number of spatial sites $N$).
One of the significant effects of $J_1$ is to introduce a macroscopic ground state
degeneracy so that the associated low temperature entropy per site is non-zero. In particular, for $J_1 \geq J_0$
 on each bond $\langle i j \rangle$ the value of $S_j$ which minimizes the energy
is not uniquely determined by $S_i$, resulting in nonvanishing ground state entropy.
Here we consider such entropy for a one dimensional chain of sites, focusing on
the case $J_1=J_0$ when, on each bond, there are three choices
$S_j = S_i, S_i \pm 1$ which have equivalent energy. There is, however, a choice of boundary
 conditions in the synthetic dimension; we begin our analysis with periodic boundary
 conditions, followed by a discussion of open boundary conditions, which is easier to achieve in experiments.
 
 With periodic boundary conditions in the synthetic dimension, the entropy for a one dimensional chain is easy to analyze
 analytically. We start with $J_1=J_0$, looking at a single site, which can take $q$ different
 values. To keep the system in its ground state, all of its neighboring sites can each only
 take three different values: identical, one more, or one less. As more and more sites
 are considered, it becomes clear that each site---other than the first---can only take
 three different values. As such, in the limit as the number of sites goes to infinity,
 the degeneracy, and therefore entropy, can be written as,
\begin{gather*}
\Omega = \lim_{N\to\infty}q\cdot3^{N-1} \\
\lim_{N\to\infty}\frac{S}{N} = \lim_{N\to\infty}\dfrac{\ln{\Omega}}{N} = \ln{3}
\end{gather*}
For $J_1>J_0$, a similar argument can be made, except each site is constrained only to take two possible values rather than three. Accordingly, in the limit as the number of sites goes to infinity, the entropy per site is found to be $\ln{2}$.

We now turn our attention to open boundary conditions in the synthetic dimension, which is
 substantially more difficult to analyze. Rather than calculating analytic expressions for
 the entropy, we establish lower and upper bounds, followed by computationally calculated
 approximations to check our results, all still for a one dimensional chain of sites. Starting with $J_1=J_0$, one can use an analogous
 argument to the previous entropy calculations. With open boundary conditions, however,
 sites are constrained to take a \textit{maximum} of three different values rather than
 \textit{strictly} three different values. It is certainly possible that a site is constrained
 such that it can only take two or even one possible value (such as when the value of a
 neighboring site is 1 or $q$). 
 Accordingly, in the limit as the number of sites goes to infinity, an upper bound to
 the ground state entropy per site is $\ln{3}$.

Still considering $J_1=J_0$, to derive a lower bound we imagine a lattice in which every
 second site takes the same value. Given that $q>2$ so that this value can be neither the maximum or minimum ($1$ or $q$), then each of the remaining sites is constrained to be either that same value, one
 more, or one less, resulting in a total of three different possibilities
 for each site. In the limit as the number of sites goes to infinity,
 the total number of microstates under this constraint and the ground
 state entropy per site can then be written as,
\begin{gather*}
\Omega \sim \lim_{N\to\infty}3^{N/2} \\
\lim_{N\to\infty}\frac{S}{N} = \lim_{N\to\infty}\dfrac{\ln{\Omega}}{N} = \dfrac{\ln{3}}{2}
\end{gather*}
Since there are certainly other possible microstates, this is a lower bound for the
 ground state entropy per site. These upper and lower bounds hold in one dimension for any system size, and also for any $q<\infty$. Similar arguments can be
 made for $J_1 > J_0$, resulting in lower and upper bounds of $\frac{\ln{2}}{2}$ and $\ln{2}$ respectively.

\begin{figure}[t]
\centering
\includegraphics[height=3.7in,width=2.5in]{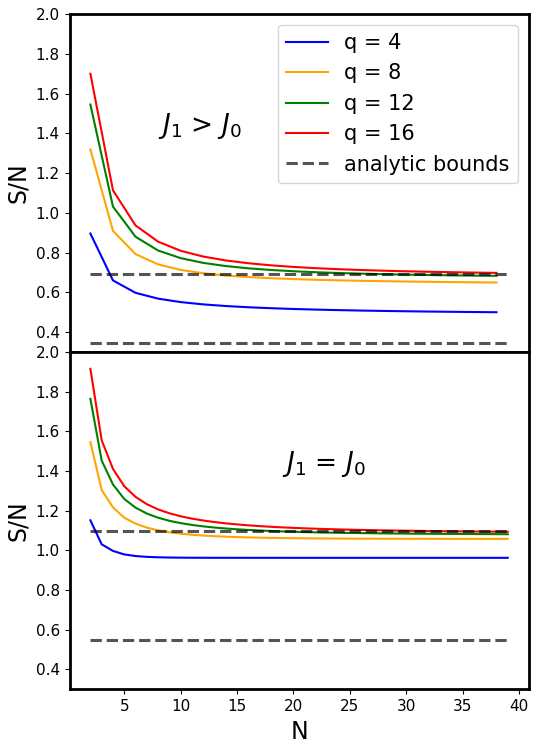}
\caption{Ground state entropy per site in one dimension for different
 $q$ values where $N$ is the number of sites in the chain. In both plots, $q=4$ is the bottom curve, followed by $q=8$ and $q=12$, finally with $q=16$ on top. As $N$ increases, the entropy decreases, and all will eventually cross below the upper bound. Larger $q$ values result in larger entropy, as expected.\label{fig:GS-entrop}
 }
\end{figure}

As mentioned previously, computational approximations
 for the ground state entropy are also calculated in one dimension for $J_1 = J_0$ and $J_1 > J_0$.
 An algorithm was created using a Pascal's triangle argument to
 count the number of microstates—and therefore entropy—for different
 chain sizes and $q$ values. This algorithm can best be understood by representing the
 one dimensional chain with a two dimensional grid of width $N$ and height $q$, where the
 horizontal axis represents the position in the chain, and the vertical
 axis being the spin value of that particular site. For $J_1 = J_0$, the number of ground states
 is exactly given by the number of ways to travel from the left side to the right side
 of the grid while only making moves directly to the right, diagonally up/right, or
 diagonally down/right. Given periodic boundary conditions in the real dimension, the
 only other constraint is that there must be at most a one value difference between
 the starting and ending height positions on the grid. For the $J_1 > J_0$ case, the
 same argument can be made, but only diagonal moves are allowed, and the ending height
 must be exactly one above or below the starting height. These types of questions,
 commonly known as pathway problems, are easily solved by
creating an augmented Pascal’s Triangle starting from the far left of the grid,
 and propagating to the right by summing previous values along all available paths. Results
 are shown in Fig.~\ref{fig:GS-entrop}. 
The best approximations are consistent with the previously established bounds.

\vskip0.08in \noindent
{\bf B. Details of ML approach}
\vskip0.1in \noindent

As mentioned previously, we used the LBC method with CNNs. In particular, our CNNs consisted of 3 separate convolution
layers in parallel with filter sizes $3\times 3$, $5\times 5$, and $7\times 7$. After raw Monte Carlo
configurations are put through each of these convolution layers, the
outputs are then concatenated back into a single tensor, where the number of neurons is equal to the total number in the concatinated output of the CNNs. This tensor is then put through two fully connected layers with ReLU activation and 512 hidden neurons, 
finally running the two output neurons through softmax to classify the 
configuration into one of the two classes. A dropout probability
 of 0.1 was used in the fully connected layers to avoid overfitting.
 
Before each training trial, our data were split into training 
(42.5\% of the data),
testing (42.5\% of the data), and validation 
(15\% of the data) splits.
The CNN is then trained using the training data split with Adam optimization, and cross-entropy loss on the validation split.
 Training ends once the
validation loss stops decreasing for five epochs, which is done to avoid 
overfitting. The initial value of the learning rate is $0.001$ and is updated with Pytorch's ReduceLROnPlateau learning
rate scheduler, which multiplies the learning rate by a factor of 0.1 with a patience of 3. Accuracy is then calculated with the test data
split, which is recorded so it can later be plotted.

Our data sets consisted of $n$ independent configurations at $t$ different temperatures,
 resulting in $n\cdot t$ total configurations. When temperatures are sufficiently
 low, however, 
 the sampling is not ergodic and the simulation spends a very long time in one
 of the degenerate ordered states.  To address this,
 `spatially translated' data were created as follows:  From each `real'
 configuration, $q-1$ `translated' configurations were generated by increasing
 the spin value on each site:  
 $S_i \rightarrow S_i +1$,
 with PBC imposed such that
 $S_i=q \rightarrow S_i=1$,
 This corresponds to moving the configuration in the synthetic dimension.
 Likewise all spatial sites are also translated a random amount $i \rightarrow i + \vec \delta$
 in the two `real' dimensions. These two steps were performed $q-1$ times,
 resulting in $q$ different configurations with the same energy as the original. 
 The final data set consists of $q \cdot n \cdot t$ total configurations. Accordingly, higher $q$ values result in larger amounts of translated data, thus increasing computation time.

\end{document}